\documentclass{llncs}
\usepackage{amsmath}
\usepackage{graphicx}
\usepackage{color}
\usepackage[utf8x]{inputenc}

\begin{document}


\title{Hop and HipHop : Multitier Web Orchestration}

\author{G{\'e}rard Berry\inst{1} and Manuel Serrano\inst{2}}
\institute{Coll{\`e}ge de France,11 place Marcelin Berthelot, 75231 Paris Cedex 05, France, Gerard.Berry@college-de-france.fr
\and 
Inria Sophia M{\'e}diterran{\'e}e, 2004 route des Lucioles, 06902 Sophia Antipolis, France, Manuel.Serrano@inria.fr}

\maketitle

\begin{abstract}
Rich applications merge classical computing, client-server
concurrency, web-based interfaces, and the complex time- and
event-based reactive programming found in embedded systems. To handle
them, we extend the Hop web programming platform by HipHop, a
domain-specific language dedicated to event-based process
orchestration. Borrowing the synchronous reactive model of Esterel,
HipHop is based on synchronous concurrency and preemption primitives
that are known to be key components for the modular design of complex
reactive behaviors. HipHop departs from Esterel by its ability to
handle the dynamicity of Web applications, thanks to the reflexivity
of Hop.  Using a music player example, we show how to modularly build
a non-trivial Hop application using HipHop orchestration code.
\end{abstract}

\section{Introduction}
\label{Introduction}

Our aim is to help programming rich applications driven by computers,
smartphones or tablets; since they interact with various external
services and devices, such applications require orchestration
techniques that merge classical computing, client-server concurrency,
web-based interfaces, and event-based programming. To achieve this, we
extend the Hop multitier web programming platform \cite{sb:cacm12} by
the new HipHop domain specific language (DSL), which is based on the
synchronous language Esterel \cite{berry2000foundations}. HipHop
orchestrates and synchronizes internal and external activities
according to timers, events generated by the network, GUIs, sensors
and devices, or internally computed conditions.

Like Esterel, Hiphop is a concurrent language based on the perfect
synchrony hypothesis: a HipHop program repeatedly reacts in conceptual
zero-delay to input events by generating output events;
synchronization and communication between parallel statements is also
performed in conceptual zero-delay. Perfect synchrony makes concurrent
programs deterministic and deadlock-free, the only non-determinism
left being that of the application environment. Its implementation is
cycle-based, execution consisting of repeated atomic cycles ``read
inputs / compute reaction / generate outputs'' in coroutine with the main Hop code.
Concurrency is
compiled away by static or dynamic sequential scheduling of
code fragments.  Cyclic execution atomicity avoids interference between
computation and input-output, which is the usual source of unexpected
non-determinism and synchronization problems for 
classical event-handler based programming.

While Esterel is limited to static applications, HipHop is designed
for dynamicity.  Its implementation on top of Hop makes it possible to
dynamically build and run orchestration programs at any time using
Hop's reflexivity facilities. It even makes it possible to modify a
HipHop program between two execution cycles (not detailed here). 
It also simplifies the language by importing Hop's data definition facilities,
expressions, modular structure, and higher-order programming
features. It relies on the Web asynchronous concurrency and messaging 
already supported by Hop.

Section \ref{Hop} briefly presents the Hop language.
Section \ref{The HipHop Programming Language} describes
HipHop and its relation with Hop. Section \ref{Applications}
presents a music application. Section \ref{Related works} briefly overviews 
related work. We conclude in Section \ref{Conclusion}.

\section{Hop}
\label{Hop}

Hop has been presented in several publications
\cite{sgl:dls06,sb:cacm12}. We only remind its essential
aspects and show some examples that should be sufficient to understand
the rest of the paper.

Hop is a Scheme-based multitier functional language. The application server-side
and client-side are both implemented within a single Hop program.
Client code is distinguished from server code by prefixing it with the
syntactic annotation `{\texttt{{\char126}}}'.  Server-side values can
be injected inside a client-side expression using a second
syntactic annotation: the `{\texttt{{\char36}}}' mark.  On the server,
the client-side code is extracted, compiled on-the-fly into standard
JavaScript, and shipped to the client. This enables Hop clients to be
executed by unmodified Web browsers.

Except for its new multitier programming style, Hop uses the standard
Web programming model. A server-side Hop program builds an HTML tree
that creates the GUI and embeds client-side code into scripts, then
ships it to the client.  AJAX-like service-based programming is made
available by service definitions, a service being a server-side
function associated with a URL.  The {\texttt{with-hop}} Hop form
triggers execution of a service.  Communication between clients and
servers is automatically performed by the Hop runtime system, with no
additional user code needed.

The Hop Web application \texttt{fib-html} below illustrates multitier
programming. It consists of a server-built Web page displaying a
three-rows table whose cells enumerate positive integers. When a cell
is clicked, the corresponding Fibonacci value is computed on the
client and displayed in a popup window. Note the
`{\texttt{{\char126}}}' signs used lines {\em{3}},{\em{6}}, {\em{7}},
and {\em{8}} which mark client-side expressions.

{\footnotesize{\begin{flushleft}
\newdimen\oldpretabcolsep
{\setlength{\oldpretabcolsep}{\tabcolsep}
\addtolength{\tabcolsep}{-\tabcolsep}
{\noindent \texttt{\begin{tabular*}{0.92\linewidth}{l@{\extracolsep{\fill}}}
\label{prog1599-1}{\scriptsize{{\textmd{{\textit{1:\,}}}}}}({\textbf{define-service}}\ (fib-html)\\
\label{prog1599-2}{\scriptsize{{\textmd{{\textit{2:\,}}}}}}\ \ \ ({\textbf{<HTML>}}\\
 \label{prog1599-3}{\scriptsize{{\textmd{{\textit{3:\,}}}}}}\ \ \ \ \ \ {\textbf{{\char126}}}({\textbf{define}}\ (fib\ x) {\textbf{\textit{;; client-side code since prefixed by {\textbf{{\char126}}}}}}\\
 \label{prog1599-4}{\scriptsize{{\textmd{{\textit{4:\,}}}}}}\ \ \ \ \ \ \ \ \ ({\textbf{if}}\ (<\ x\ 2)\ 1\ (+\ (fib\ (-\ x\ 1))\ (fib\ (-\ x\ 2)))))\\
 \label{prog1599-7}{\scriptsize{{\textmd{{\textit{5:\,}}}}}}\ \ \ \ \ \ ({\textbf{<TABLE>}}\\
 \label{prog1599-8}{\scriptsize{{\textmd{{\textit{6:\,}}}}}}\ \ \ \ \ \ \ \ \ ({\textbf{<TR>}}\ ({\textbf{<TD>}}\ "fib(1)"\ :onclick\ {\textbf{{\char126}}}(alert\ (fib\ 1))))\\
 \label{prog1599-9}{\scriptsize{{\textmd{{\textit{7:\,}}}}}}\ \ \ \ \ \ \ \ \ ({\textbf{<TR>}}\ ({\textbf{<TD>}}\ "fib(2)"\ :onclick\ {\textbf{{\char126}}}(alert\ (fib\ 2))))\\
 \label{prog1599-10}{\scriptsize{{\textmd{{\textit{8:\,}}}}}}\ \ \ \ \ \ \ \ \ ({\textbf{<TR>}}\ ({\textbf{<TD>}}\ "fib(3)"\ :onclick\ {\textbf{{\char126}}}(alert\ (fib\ 3)))))))\\
 \end{tabular*} }} \setlength{\tabcolsep}{\oldpretabcolsep}
}\end{flushleft} }}
\noindent Let us modify the example to illustrate some Hop niceties.
Instead of building the rows by hand, we let Hop compute
them. The new Hop program uses the {\texttt{(iota 3)}}
expression (line {\em{9}}) that evaluates to the list (1, 2, 3) and the
{\texttt{map}} functional operator that applies a
function to all the elements of a list. The {\texttt{{\char36}i}} expression (line {\em{8}}) denotes the value
of {\texttt{i}} on the server at HTML document elaboration time, seamlessly 
exported to the client code:

{\footnotesize{\begin{flushleft}
\newdimen\oldpretabcolsep
{\setlength{\oldpretabcolsep}{\tabcolsep}
\addtolength{\tabcolsep}{-\tabcolsep}
{\noindent \texttt{\begin{tabular*}{0.92\linewidth}{l@{\extracolsep{\fill}}}
\label{prog1636-1}{\scriptsize{{\textmd{{\textit{1:\,}}}}}}({\textbf{define-service}}\ (fib-html)\\
 \label{prog1636-2}{\scriptsize{{\textmd{{\textit{2:\,}}}}}}\ \ \ ({\textbf{<HTML>}}\\
 \label{prog1636-3}{\scriptsize{{\textmd{{\textit{3:\,}}}}}}\ \ \ \ \ \ {\textbf{{\char126}}}({\textbf{define}}\ (fib\ x)\ ...)\\
 \label{prog1636-4}{\scriptsize{{\textmd{{\textit{4:\,}}}}}}\ \ \ \ \ \ ({\textbf{<TABLE>}}\\
 \label{prog1636-5}{\scriptsize{{\textmd{{\textit{5:\,}}}}}}\ \ \ \ \ \ \ \ \ (map\ ({\textbf{lambda}}\ (i)\\
 \label{prog1636-6}{\scriptsize{{\textmd{{\textit{6:\,}}}}}}\ \ \ \ \ \ \ \ \ \ \ \ \ \ \ \ \ ({\textbf{<TR>}}\\
 \label{prog1636-7}{\scriptsize{{\textmd{{\textit{7:\,}}}}}}\ \ \ \ \ \ \ \ \ \ \ \ \ \ \ \ \ \ \ \ ({\textbf{<TD>}}\ "fib("\ i\ ")"\\
 \label{prog1636-8}{\scriptsize{{\textmd{{\textit{8:\,}}}}}}\ \ \ \ \ \ \ \ \ \ \ \ \ \ \ \ \ \ \ \ \ \ \ :onclick\ {\textbf{{\char126}}}(alert\ (fib\ {\textbf{{\char36}}}i)))))\\
 \label{prog1636-9}{\scriptsize{{\textmd{{\textit{9:\,}}}}}}\ \ \ \ \ \ \ \ \ \ \ \ (iota\ 3)))))\\
 \end{tabular*} }} \setlength{\tabcolsep}{\oldpretabcolsep}
}\end{flushleft}
}}
\noindent Before delivery to a client, the server-side document is compiled on
the server into regular HTML and JavaScript. It can then be executed
by all standard browsers.

\section{The HipHop Programming Language}
\label{The HipHop Programming Language}

HipHop embeds the reactive primitives of Esterel
\cite{berry2000foundations} within Hop while making maximal usage of
Hop's expressive power.  By convention, the `{\texttt{{\char38}}}'
suffix is associated with HipHop code.  Technically speaking, a HipHop
form should be seen in two ways. First, it is a Hop constructor that
builds a Hop value that represents a HipHop abstract syntax node. This
makes it possible to dynamically build and run HipHop programs from
within Hop. Second, it is a temporal statement executed by a {\em
  reactive machine} that communicates with Hop using logical HipHop
events built by Hop out of physical or programmed events.

The reactive machine is triggered by Hop and perform conceptually
instantaneous and deterministic {\em reactions} to its input HipHop events, generating output HipHop events.

\subsection{HipHop Events}
\label{HipHop events}

HipHop logical events are abstract Hop values of class {\texttt{HipHopEvent}}.
They can be inputs and outputs of the reactive
machine or local to the HipHop program, then helping
synchronization and communication between its concurrent parts.
HipHop events have an optional boolean presence/absence {\em{status}}
and an optional {\em data value}. The
status and value of each event are unique in each reaction and
broadcast to the parallel components of the HipHop program.

The status of an event is {\em{absent}} by defaut. Input events are
set {\em{present}} from Hop prior to the reaction using the
{\texttt{hiphop-input!}} Hop form; this determines the {\em{input
context}}. Local and output events are set present from within the
HipHop program by executing the {\texttt{emit{\char38}}} statement.
The status of an event $e$ is not memorized between successive reactions. It is read using the {\texttt{(now{\char38} $e$)}} form, while
the status at the previous reaction is read using the {\texttt{(pre{\char38} $e$)}}
form.

The data value of an event is defined when setting the status, either
from Hop using {\texttt{hiphop-input!}} for an input or by
{\texttt{emit{\char38}}} for an output or local. Contrarily to the
status, the value is memorized between reactions. The current value of
$e$ is returned by the {\texttt{(val{\char38} $e$)}} form, while the
value at the previous reaction is read using the
{\texttt{(preval{\char38} $e$)}} form. As for Esterel, several
emissions can occur for the same event in the same reaction; they are
said to be {\em{simultaneous}}.  In that case, the final value of the
event is obtained by combining the individually emitted values using a
combination function specified in the event Hop object declaration.

\subsection{Reactive Machines and their Reactions}
\label{Reactive Machines and their Reactions}

Reactive machines interface Hop and HipHop. A machine $M$
is defined by its HipHop input/output logical event interface and its
HipHop program. 

Hop delivers an input event {\texttt{A}} with value $v$ to a reactive machine $M$ 
using the form {\texttt{(hiphop-input! $M$ A $v$)}}. Any number of inputs can be delivered
before a reaction; they are only valid for this reaction.
A reaction is triggered from within Hop by
{\texttt{(hiphop-react! $M$)}}. Determining when a machine should react is
solely Hop's responsibility. However, to simplify a common case,
it is possible to write {\texttt{(hiphop-input-and-react! $M$ A
$v$)}} to pass an input and trigger a reaction right away.

A reaction may trigger output events, the actual output action being
performed by associated Hop listeners associated with the events and
stored in the reactive machine. To handle data, a reaction may also
trigger the evaluation of Hop expressions using the
{\texttt{atom{\char38}}} HipHop statement, see Section \ref{HipHop core
  statements}.

Seen from Hop, a HipHop reaction is simply a standard function
call. Seen from HipHop, the execution of the reaction is conceptually
performed in zero-delay, the HipHop program sleeping between two
successive reactions and remembering its control state from one reaction to the next.
This coroutine execution
scheme avoids interference between input event registering and
reactions, which is a common cause of unwanted non-determinism and
deadlocks with classical threading techniques.

A reactive machine can be executed on the server or
shipped to and executed on a client, because it is a standard Hop
object. Several reactive machines can coexist in the
same application, making it possible to use a GALS programming model
(Globally Asynchronous, Locally Synchronous) without extra overhead.
This will not be detailed here.

\subsection{HipHop Intuitive Execution Semantics}
\label{HipHop Intuitive Execution Semantics}

The reactive code is based on deterministic sequencing,
concurrency, and temporal statements inspired from Esterel
\cite{berry2000foundations}. Control positions are memorized from one
reaction to the next. To illustrate sequencing, consider the following sequence:

{\footnotesize{\begin{flushleft}
\newdimen\oldpretabcolsep
{\setlength{\oldpretabcolsep}{\tabcolsep}
\addtolength{\tabcolsep}{-\tabcolsep}
{\noindent \texttt{\begin{tabular*}{0.90\linewidth}{l@{\extracolsep{\fill}}}
\ \ \ (seq{\char38}\\
\ \ \ \ \ \ (await{\char38}\ A)\\
\ \ \ \ \ \ (await{\char38}\ B)\\
\ \ \ \ \ \ (emit{\char38}\ O)\end{tabular*}
}}
\setlength{\tabcolsep}{\oldpretabcolsep}
}\end{flushleft}
}}
\noindent where {\texttt{A}} and {\texttt{B}} (resp. {\texttt{O}}) are input
(resp. output) HipHop events. Intuitively, the code waits for {\texttt{A}} and then {\texttt{B}}
to be present, before emitting {\texttt{O}}
and terminating synchronously: {\texttt{O}} is emitted within
the reaction triggered by {\texttt{B}}. Technically, at first reaction, the HipHop
control flow stops on {\texttt{(await{\char38} A)}}, and yields back
control to Hop. HipHop control stays there at each subsequent reaction until
the first reaction where {\texttt{A}} is present. In this reaction,
control immediately moves to {\texttt{(await{\char38} B)}} and stays
there until the next reaction where {\texttt{B}} is present. During
this reaction, and without further delay, it outputs {\texttt{O}} and
terminates.

To illustrate concurrency, consider now the following HipHop code:

{\footnotesize{\begin{flushleft}
\newdimen\oldpretabcolsep
{\setlength{\oldpretabcolsep}{\tabcolsep}
\addtolength{\tabcolsep}{-\tabcolsep}
{\noindent \texttt{\begin{tabular*}{0.90\linewidth}{l@{\extracolsep{\fill}}}
\ \ \ (seq{\char38}\\
\ \ \ \ \ \ (par{\char38}\ (await{\char38}\ A)\ (await{\char38}\ B))\\
\ \ \ \ \ \ (emit{\char38}\ O)\end{tabular*}
}}
\setlength{\tabcolsep}{\oldpretabcolsep}
}\end{flushleft}
}}
\noindent Here, {\texttt{A}} and {\texttt{B}} are waited for in parallel, not in
sequence. The \texttt{par{\char38}} statement
terminates when all its arms are terminated. Thus, {\texttt{O}} is
emitted exactly when the last of {\texttt{A}} and {\texttt{B}}
occurs (note that {\texttt{A}} and {\texttt{B}} may be both present
in the same reaction if they have been both input into the reactive machine
before the reaction is triggered). In HipHop, concurrency and
sequencing can be mixed arbitrarily, and the same holds for all other
instructions.

We say that a statement that starts and terminates in the same
reaction is {\em{instantaneous}} or {\em{immediate}}; this is the case for
{\texttt{emit{\char38}}}.  Otherwise, we say that the statement {\em{pauses}}, waiting
for the next reaction, and we call it a {\em{delay statement}}; this is
the case for {\texttt{await{\char38}}}.  Things that happen in the same reaction are
called {\em{simultaneous}}. This is of course a conceptual notion in
terms of abstract reactions, not a physical one.

\subsection{HipHop Core Statements}
\label{HipHop core statements}

As for Esterel, statements are divided into {\em{core statements}},
which are primitives and handy {\em{derived statements}}. Thanks to
Hop's reflexivity, derived statements can be trivially defined from
core statements using Hop. We first detail the core statements.

The {\texttt{nothing{\char38}}} statement does nothing and terminates
instantaneously. It is the HipHop no-op. 
The {\texttt{(emit{\char38} $e$ }}[$v$]\texttt{)} statement emits its event $e$ with value
determined by optional
$v$. It terminates instantaneously.
The {\texttt{(atom{\char38} $expr$)}} statement calls Hop to executes the $expr$ Hop
expression; it is instantaneous, which means that its Hop argument
execution time should be kept negligible in practice.

The {\texttt{pause{\char38}}} statement delays execution by one reaction:
it pauses for the reaction and terminates at the next
reaction.

The {\texttt{(if{\char38} $test$ $then$ $else$)}} statement instantaneously evaluates $test$. If the
result is true, it immediately starts $then$ and
behaves as it from then on; otherwise, it does the same with $else$. These can be arbitrary 
HipHop statements. Termination of the {\texttt{if{\char38}}} statement is
instantaneously triggered by termination of the selected branch.
The {\texttt{seq{\char38}}} statement executes its arguments in order: the first
one starts immediately when the sequence starts; when it
terminates, be it immediately or in a delayed way, the second argument
is immediately started, etc. For instance, {\texttt{(seq{\char38} (emit{\char38} A) (emit{\char38}
B))}} immediately emits {\texttt{A}} and {\texttt{B}}, which are seen as
simultaneous within the reaction, while {\texttt{(seq{\char38} (emit{\char38} A) (pause{\char38})
(emit{\char38} B))}} emits {\texttt{A}} and {\texttt{B}} in two successive reactions.

The {\texttt{loop{\char38}}} statement is a loop-forever, equivalent to the
infinite sequential repetition of its argument statements, themselves
implicitly evaluated in sequence.  For instance, {\texttt{(loop{\char38} (pause{\char38})
(emit{\char38} A))}} waits for the next reaction and then keeps emitting {\texttt{A}}
at each reaction. 
Exiting a {\texttt{loop{\char38}}} can only be
done by using the {\texttt{trap{\char38}}}/{\texttt{exit{\char38}}}, {\texttt{abort{\char38}}}, and {\texttt{until{\char38}}} statements, see below.

The {\texttt{par{\char38}}} statement starts its arguments
concurrently and terminates when the last of them
terminates. Therefore, {\texttt{(par{\char38} (await{\char38} A)
(await{\char38} B))}} immediately terminates when both {\texttt{A}} and
{\texttt{B}} have been received. Remember that all arms of a
{\texttt{par{\char38}}} statement see all statuses and values of all
(visible) events in exactly the same way.

The {\texttt{suspend{\char38}}} statement immediately starts its
body. At all following instants, it suspends (freezes) the execution
of its body for the reaction when its condition is true. The
{\texttt{suspend{\char38}}} statement terminates if its body is
executed and terminates. For instance,

{\footnotesize{\begin{flushleft}
\newdimen\oldpretabcolsep
{\setlength{\oldpretabcolsep}{\tabcolsep}
\addtolength{\tabcolsep}{-\tabcolsep}
{\noindent \texttt{\begin{tabular*}{0.90\linewidth}{l@{\extracolsep{\fill}}}
\ \ (suspend{\char38}\ (now{\char38}\ A)\\
\ \ \ \ \ (loop{\char38}\\
\ \ \ \ \ \ \ \ (emit{\char38}\ B)\\
\ \ \ \ \ \ \ \ (pause{\char38}))\end{tabular*}
}}
\setlength{\tabcolsep}{\oldpretabcolsep}
}\end{flushleft}
}}
\noindent emits {\texttt{B}} at first instant and at all subsequent instants where {\texttt{A}}
is absent.

The {\texttt{trap{\char38}}} statement defines a named exit point for its body.
The {\texttt{exit{\char38}}} statement provokes immediate termination of the
corresponding {\texttt{trap{\char38}}} statement, as well as immediate termination of
all concurrent statements within the {\texttt{trap{\char38}}} body, which do normally
receive the control at that instant.

The {\texttt{local{\char38}}} statement declares local events in the
first argument list. Their scope is the body, which is the implicitly
{\texttt{seq{\char38}}} list of the remaining HipHop arguments. The
declared events are not visible from Hop. A {\texttt{local{\char38}}}
statement terminates when its body does.

\subsection{HipHop Derived Statements}
\label{HipHop derived statements}

The derived statements can be easily defined from the kernel ones using Hop.
The {\texttt{halt{\char38}}} statement 
pauses forever; it is defined as \texttt{(loop{\char38} (pause{\char38}))}.
The {\texttt{sustain{\char38}}} statement
keeps emitting its event at each reaction. The {\texttt{await{\char38}}} statement pauses and 
waits for
its expression to become true and terminates:

{\footnotesize{\begin{flushleft}
\newdimen\oldpretabcolsep
{\setlength{\oldpretabcolsep}{\tabcolsep}
\addtolength{\tabcolsep}{-\tabcolsep}
{\noindent \texttt{\begin{tabular*}{0.90\linewidth}{l@{\extracolsep{\fill}}}
(define\ (await{\char38}\ evt)\\
\ \ \ (trap{\char38}\ (done)\\
\ \ \ \ \ \ (loop{\char38}\\
\ \ \ \ \ \ \ \ \ (pause{\char38})\\
\ \ \ \ \ \ \ \ \ (if{\char38}\ (now{\char38}\ evt)\ (exit{\char38}\ done)))))\\
\end{tabular*}
}}
\setlength{\tabcolsep}{\oldpretabcolsep}
}\end{flushleft}
}}
\noindent The {\texttt{abort{\char38}}} statement
instantaneously kills its sequential body when its condition becomes true, not
passing the control to its body in this reaction; this is what we call
{\em{strong abortion}}:

{\footnotesize{\begin{flushleft}
\newdimen\oldpretabcolsep
{\setlength{\oldpretabcolsep}{\tabcolsep}
\addtolength{\tabcolsep}{-\tabcolsep}
{\noindent \texttt{\begin{tabular*}{0.90\linewidth}{l@{\extracolsep{\fill}}}
(define\ (abort{\char38}\ evt\ .\ stmt-list)\\
\ \ \ (trap{\char38}\ (done)\\
\ \ \ \ \ \ (par{\char38}\\
\ \ \ \ \ \ \ \ \ (suspend{\char38}\ (now{\char38}\ evt)\ stmt-list)\\
\ \ \ \ \ \ \ \ \ (await{\char38} evt\ (exit{\char38}\ done)))))\\
\end{tabular*}
}}
\setlength{\tabcolsep}{\oldpretabcolsep}
}\end{flushleft}
}}
\noindent
The {\texttt{until{\char38}}} statement instantaneously kills its body
when its condition becomes true, but only at the end of the reaction,
passing the control to its body for the last time at that reaction as
for an exited {\texttt{trap{\char38}}}; this is what we call {\em{weak
    abortion}}. The {\texttt{loop-each{\char38}}} statement
immediately starts its body, and then stongly kills it and restarts it
immediately whenever its condition becomes true.  The
{\texttt{every{\char38}}} statement is similar but starts by waiting
for the condition instead of immediately starting its body (see
\cite{CompilingEsterel}).

Once defined, these statements can be freely used in HipHop
programs. Note that this makes the language fully user-extensible. One
can also build dynamically statements from dynamic values computed
during Hop execution, and even dynamically modify the program between
two reactions, for instance to use and orchestrate services dynamically detected at runtime.
Note also that there is no need to redefine the
basic arithmetic, list, and string expressions since Hop's ones can be
reused (with some care however, no details given here). 

\section{An Application Example}
\label{Applications}

\begin{figure}[htb]
\begin{center}\includegraphics[width=0.58\linewidth]{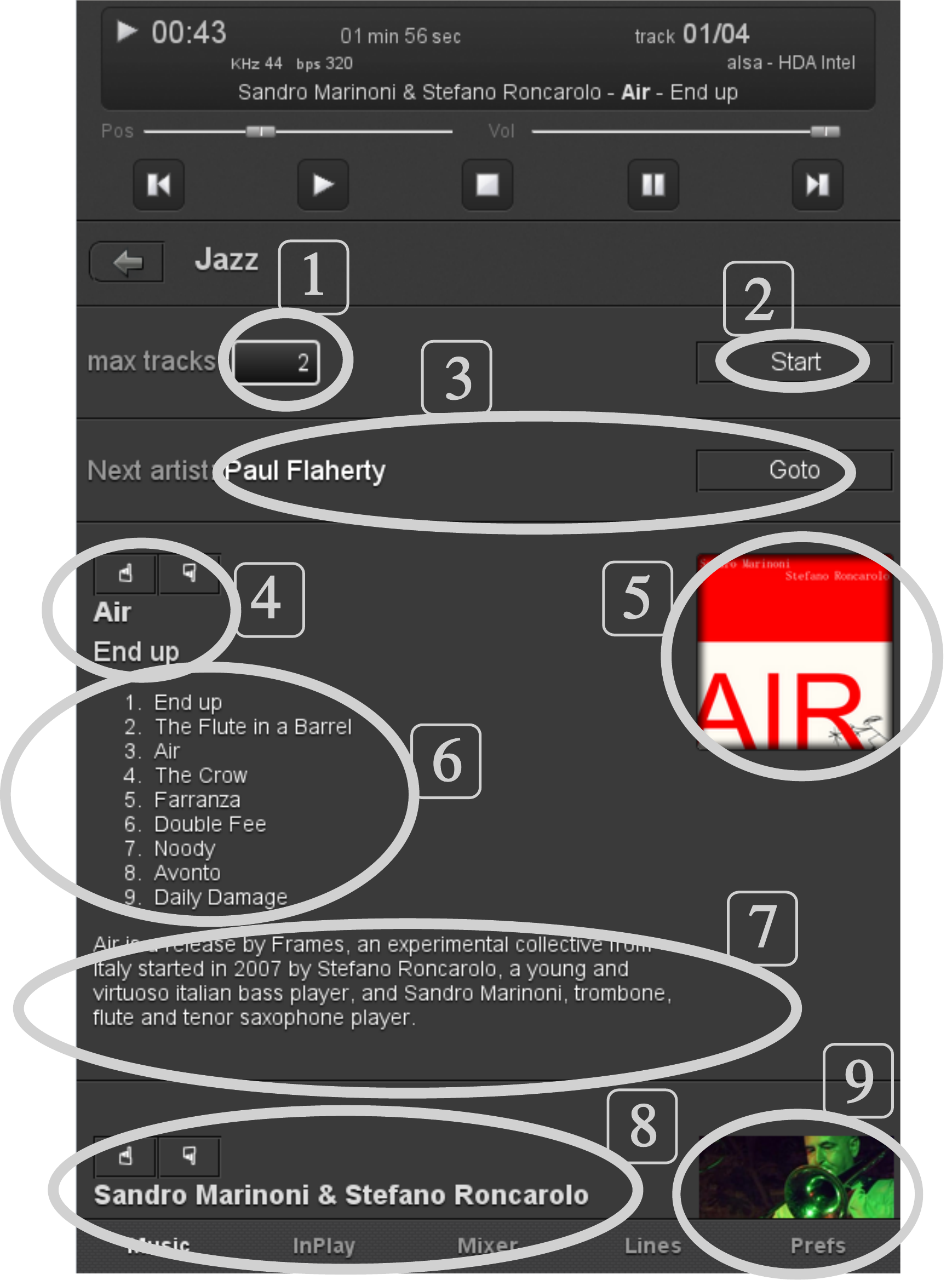}\end{center} 

\vspace{1ex}
\caption{The smartphone screenshot of HopFM. Ellipses
show active zones managed by Hiphop.  Zone 1 sets
the number of tracks to play per artist.  Zone 2 initiates
the music play by starting the reactive
machine. Zone 3 displays the name of the next artist and the
button to switch to his music. Zone 4 reports
the current track   name and the album name; 
the two top buttons enable the user to emit
positive or negative recommendations.  Zone 5 displays the album
image, zone 6 the album's track list, zone 7
comments about the album. Zone 8 displays the artist name and
two buttons to emit positive or negative recommendations.  Zone
9 displays an image of the artist.
\label{hophifi-fig}}\end{figure}

We build a Lastfm-like smart music player called HopFM that orchestrates
musical content and related information available on third
party Web sites. It plays music continuously, switching from one
artist to another according to musical similarities. It automatically
fetches and displays information about music and authors.

The user first selects a musical genre in a dynamically discovered
list.  This activates the Hop control screen of Figure
\ref{hophifi-fig}. The top of screen is used to adjust the volume,
pause the music, switch to the next track, etc. Below stands the
{\textbf{{\footnotesize{{\footnotesize{\fbox{\sf{Start}}}}}}}} button
(zone 2, Figure \ref{hophifi-fig}) that starts HopFM when
clicked. HopFM then searches the internet for a random artist of the
selected genre, downloads a limited number of tracks of this artist,
and starts playing. When a new track starts, HopFM looks for related
information on the internet: the associated album, its image, its
track list, and reviews.  In parallel, HopFM searches and displays
information about the played artist: his biography, his discography,
some news, and some blogs and reviews.

Also in parallel, the application searches the Web for similar artists
to be played later on, either when the currently downloaded tracks are
over or when the user clicks the
{\textbf{{\footnotesize{{\footnotesize{\fbox{\sf{Goto}}}}}}}} button
(zone 3). If no similar artist is
found, or if all the known similar artists have been played, HopFM
randomly chooses a new random artist of the genre and starts again.
HopFM keeps running until the user clicks the top stop button.

The music player uses the following third-party services to retrieve
musical contents and information about the tracks and artists:
{\em Free Music Archive}\footnote{{\texttt{http://freemusicarchive.org}}.}
(FMA), which
provides mp3 music and informations about the tracks, albums, and
artists;
{\em EchoNest}\footnote{{\texttt{http://echonest.com}}.}, a
database about tracks, artists, images, artist similarities, and alignment information
between other databases; 
{\em MusicBrainz}\footnote{{\texttt{http://musicbrainz.org}}.}, an open music encyclopedia
 that collects music metadata.

HipHop orchestrates the requests to these remote services. For
instance, to search for an artist image, several requests to several
sites are emitted simultaneously. As soon as the first one completes,
the other ones are aborted and the GUI is updated. If the artist
changes before any request completes, all current requests are
aborted.

HipHop also handles user interactions and signals raised by
the actual hardware music player. Web services returns or errors, user
interaction, and music player events are the three sources of
external asynchrony. They are all managed
uniformly by HipHop, local synchronous parallelism and communication being
essential to regulate and orchestrate external asynchrony.

\subsection{HopFM Implementation}
\label{HopFM}

Clicking
{\textbf{{\footnotesize{{\footnotesize{\fbox{\sf{Start}}}}}}}} invokes
the Hop client-side {\texttt{hopfm-play}} function, which
constructs a reactive HipHop machine, creates various events, binds
the external Hop events to the HipHop interface events,
creates a HipHop reactive program, loads it into the machine, and
eventually triggers the first reaction:

{\footnotesize{\begin{flushleft}
\newdimen\oldpretabcolsep
{\setlength{\oldpretabcolsep}{\tabcolsep}
\addtolength{\tabcolsep}{-\tabcolsep}
{\noindent \texttt{\begin{tabular*}{0.90\linewidth}{l@{\extracolsep{\fill}}}
({\textbf{define}}\ (hopfm-play\ catalog\ genre{\textbf{::genre}})\\
\ \ \ \\
\ \ \ ({\textbf{define}}\ musicstate\ ({\textbf{instantiate}}{\textbf{::HipHopEvent}}))\\
\ \ \ ({\textbf{define}}\ track\ ({\textbf{instantiate}}{\textbf{::HipHopEvent}}))\\
\ \ \ ({\textbf{define}}\ artist\ ({\textbf{instantiate}}{\textbf{::HipHopEvent}}))\\
\ \ \ ({\textbf{define}}\ playlist\ ({\textbf{instantiate}}{\textbf{::HipHopEvent}}))\\
\ \ \ ...\\
\\
\ \ \ ({\textbf{define}}\ ({\textbf{hopfm{\char38}}})\ ...)\\
\ \ \ \\
\ \ \ ({\textbf{let}}\ ((M\ ({\textbf{instantiate}}{\textbf{::HopHifiMachine}}\\
\ \ \ \ \ \ \ \ \ \ \ \ \ \ \ (program\ ({\textbf{hopfm{\char38}}})))))\\
\ \ \ \ \ \ {\textbf{{\textit{;; bind the machine to external HopHifi events}}}}\\
\ \ \ \ \ \ (add-event-listeners!\ M\ musicstate\ track\ artist)\\
\ \ \ \ \ \ {\textbf{{\textit{;; trigger first HipHop reaction}}}}\\
\ \ \ \ \ \ (hiphop-react!\ M)))\\
\end{tabular*}
}}
\setlength{\tabcolsep}{\oldpretabcolsep}
}\end{flushleft}
}}
\noindent The Hop function {\texttt{add-event-listeners!}} connects
the actual hardware player to the HipHop program. It binds
Hop listeners that forward the events to HipHop.

{\footnotesize{\begin{flushleft}
\newdimen\oldpretabcolsep
{\setlength{\oldpretabcolsep}{\tabcolsep}
\addtolength{\tabcolsep}{-\tabcolsep}
{\noindent \texttt{\begin{tabular*}{0.90\linewidth}{l@{\extracolsep{\fill}}}
({\textbf{define}}\ (add-event-listeners!\ M\ musicstate\ track\ artist)\\
\ \ \ ({\textbf{add-event-listener!}}\ server\ "hophifi-state"\\
\ \ \ \ \ \ {\textbf{{\textit{;; listener called when the music player state changes}}}}\\
\ \ \ \ \ \ ({\textbf{lambda}}\ (e)\\
\ \ \ \ \ \ \ \ \ {\textbf{{\textit{;; forward the Hop event to HipHop}}}}\\
\ \ \ \ \ \ \ \ \ (hiphop-input-and-react!\ M musicstate\ (event-value\ e))))\\
\ \ \ ({\textbf{add-event-listener!}}\ server\ "hophifi-track"\\
\ \ \ \ \ \ {\textbf{{\textit{;; listener is called when a new track starts}}}}\\
\ \ \ \ \ \ {\textbf{{\textit{;; or when the playlist changes}}}}\\
\ \ \ \ \ \ ({\textbf{lambda}}\ (e)\\
\ \ \ \ \ \ \ \ \ ({\textbf{let*}}\ ((ev\ (event-value\ e))\\
\ \ \ \ \ \ \ \ \ \ \ \ \ \ \ \ (tk\ (list-ref\ ev.playlist\ ev.song)))\\
\ \ \ \ \ \ \ \ \ \ \ \ {\textbf{{\textit{;; forward the track and artist to HipHop}}}}\\
\ \ \ \ \ \ \ \ \ \ \ \ (hiphop-input!\ M track\ tk)\\
\ \ \ \ \ \ \ \ \ \ \ \ (hiphop-input!\ M artist\ (track-artist tk))\\
\ \ \ \ \ \ \ \ \ \ \ \ (hiphop-react!\ M)))))\\
\end{tabular*}
}}
\setlength{\tabcolsep}{\oldpretabcolsep}
}\end{flushleft}
}}
\noindent The HipHop program \texttt{hopfm{\char38}} runs a number of
components in synchronous deterministic parallel, each in charge of a
specific task. These components synchronize each other by
communicating synchronously using the HipHop events defined above:
\texttt{track}, \texttt{artist}, \texttt{playlist}, \texttt{album},
etc.  The {\texttt{random-playlist{\char38}}} component looks for
random playlists, each playlist being associated with an artist;
{\texttt{playlist{\char38}}} waits for playlist changes and starts
searching for the next artist; {\texttt{track{\char38}}} waits for the
hardware player to start a new track, checks if it belongs to a
different album or to a different artist and, in this case, emits the
HipHop events \texttt{album} and \texttt{artist} towards the other
components; {\texttt{artist-info{\char38}}} waits for a new artist
event, searches the internet for information about that artist, and
emits an event that lets the {\texttt{gui{\char38}}} component update
the screen.  The HipHop program stops when the user clicks the main
{\textbf{{\footnotesize{{\footnotesize{\fbox{\sf{stop}}}}}}}} button,
which raises the {\texttt{musicstate}} HipHop event with value
{\texttt{stop}}; the enclosing \texttt{until{\char38}} statement then
generates a global preemption that kills all internal activities and
terminates; termination can also occur if no artist is found
(\texttt{musicstate} value \texttt{ended}).

{\footnotesize{\begin{flushleft}
\newdimen\oldpretabcolsep
{\setlength{\oldpretabcolsep}{\tabcolsep}
\addtolength{\tabcolsep}{-\tabcolsep}
{\noindent \texttt{\begin{tabular*}{0.90\linewidth}{l@{\extracolsep{\fill}}}
({\textbf{define}}\ ({\textbf{hopfm{\char38}}})\\
\ \ \ ({\textbf{until{\char38}}}\ (memq\ ({\textbf{val{\char38}}}\ musicstate)\ '(stop\ ended))\\
\ \ \ \ \ \ ({\textbf{par{\char38}} {\textit{;; running all the components in synchronous parallel}}}\\
\ \ \ \ \ \ \ \ \ {\textbf{{\textit{;; peek a random playlist}}}}\\
\ \ \ \ \ \ \ \ \ ({\textbf{random-playlist{\char38}}}\ catalog\ genre\ playlist)\\
\ \ \ \ \ \ \ \ \ {\textbf{{\textit{;; playlist manager}}}}\\
\ \ \ \ \ \ \ \ \ ({\textbf{playlist{\char38}}}\ playlist)\\
\ \ \ \ \ \ \ \ \ {\textbf{{\textit{;; deal with new tracks}}}}\\
\ \ \ \ \ \ \ \ \ ({\textbf{track{\char38}}}\ track\ album\ artist)\\
\ \ \ \ \ \ \ \ \ {\textbf{{\textit{;; manage new artists}}}}\\
\ \ \ \ \ \ \ \ \ ({\textbf{artist-info{\char38}}}\ catalog\ genre\ artist\ bio\ discog\ similar\ playlist)\\
\ \ \ \ \ \ \ \ \ {\textbf{{\textit{;; update the gui}}}}\\
\ \ \ \ \ \ \ \ \ ({\textbf{gui{\char38}}}\ musicstate\ track\ album\ artist\ bio\ discog\ similar))))\\
\end{tabular*}
}}
\setlength{\tabcolsep}{\oldpretabcolsep}
}\end{flushleft}
}}
\noindent Let us detail {\texttt{random-playlist{\char38}}}. Its
operation requires two steps: calling FMA for a random artist of the
desired genre and checking that this artist has published music. The
FMA request is proxied via the Hop server using the HipHop
{\texttt{with-hop{\char38}}} statement that takes as parameter a
service call and a HipHop event to emit if the request completes
successfully; {\texttt{with-hop{\char38}}} simply terminates silently
otherwise. Note that the artist found is kept local, since the global
artist handled by other modules is the one currently played.

{\footnotesize{\begin{flushleft}
\newdimen\oldpretabcolsep
{\setlength{\oldpretabcolsep}{\tabcolsep}
\addtolength{\tabcolsep}{-\tabcolsep}
{\noindent \texttt{\begin{tabular*}{0.90\linewidth}{l@{\extracolsep{\fill}}}
({\textbf{define}}\ ({\textbf{random-playlist{\char38}}}\ catalog\ genre\ playlist)\\
\ \ \ ({\textbf{trap{\char38}}}\ found\\
\ \ \ \ \ \ {\textbf{{\textit{;; start looping}}}}\\
\ \ \ \ \ \ ({\textbf{loop{\char38}}}\\
\ \ \ \ \ \ \ \ \ {\textbf{{\textit{;; creates two local events }}}}\\
\ \ \ \ \ \ \ \ \ ({\textbf{local{\char38}}}\ ((local-artist\ ({\textbf{instantiate}}{\textbf{::HipHopEvent}}))\\
\ \ \ \ \ \ \ \ \ \ \ \ \ \ \ \ \ \ (local-playlist\ ({\textbf{instantiate}}{\textbf{::HipHopEvent}})))\\
\ \ \ \ \ \ \ \ \ \ \ \ {\textbf{{\textit{;; get a random artist from FMA}}}}\\
\ \ \ \ \ \ \ \ \ \ \ \ ({\textbf{with-hop{\char38}}}\ ({\textbf{{\char36}}}hopfm/genre/artist/random\ genre\ catalog)\\
\ \ \ \ \ \ \ \ \ \ \ \ \ \ \ local-artist)\\
\ \ \ \ \ \ \ \ \ \ \ \ {\textbf{{\textit{;; get the tracks of that artist}}}}\\
\ \ \ \ \ \ \ \ \ \ \ \ ({\textbf{with-hop{\char38}}}\ ({\textbf{{\char36}}}hopfm/artist/tracks\ ({\textbf{val{\char38}}}\ local-artist))\\
\ \ \ \ \ \ \ \ \ \ \ \ \ \ \ local-playlist)\\
\ \ \ \ \ \ \ \ \ \ \ \ ({\textbf{if{\char38}}}\ (pair?\ ({\textbf{val{\char38}}}\ local-playlist))\\
\ \ \ \ \ \ \ \ \ \ \ \ \ \ \ {\textbf{{\textit{;; an artist with music is found}}}}\\
\ \ \ \ \ \ \ \ \ \ \ \ \ \ \ ({\textbf{seq{\char38}}}\\
\ \ \ \ \ \ \ \ \ \ \ \ \ \ \ \ \ \ ({\textbf{emit{\char38}}}\ playlist\ ({\textbf{val{\char38}}}\ playlist))\\
\ \ \ \ \ \ \ \ \ \ \ \ \ \ \ \ \ \ ({\textbf{exit{\char38}}}\ found)))))))\\
\end{tabular*}
}}
\setlength{\tabcolsep}{\oldpretabcolsep}
}\end{flushleft}
}}
\noindent The {\texttt{artist-info{\char38}}} component searches in parallel an image
of the current artist, information about that artist, and a similar
artist with a playlist. it outputs \texttt{bio}, \texttt{discog},
\texttt{similar}, and \texttt{playlist} towards the other components
as soon as the corresponding information has been found.

{\footnotesize{\begin{flushleft}
\newdimen\oldpretabcolsep
{\setlength{\oldpretabcolsep}{\tabcolsep}
\addtolength{\tabcolsep}{-\tabcolsep}
{\noindent \texttt{\begin{tabular*}{0.90\linewidth}{l@{\extracolsep{\fill}}}
({\textbf{define}}\ ({\textbf{artist-info{\char38}}}\ catalog\ genre\ artist\ bio\ discog\ similar\ playlist)\\
\ \ \ ({\textbf{every{\char38}}}\ ({\textbf{now{\char38}}}\ artist)\ {\textbf{{\textit{;; we have a playlist for that artist}}}}\\
\ \ \ \ \ \ ({\textbf{par{\char38}}}\\
\ \ \ \ \ \ \ \ \ {\textbf{{\textit{;; request a similar artist list}}}}\\
\ \ \ \ \ \ \ \ \ ({\textbf{similar-artist{\char38}}}\ catalog\ genre\ playlist\ artist\ similar)\\
\ \ \ \ \ \ \ \ \ {\textbf{{\textit{;; fetch artist biography and discography}}}}\\
\ \ \ \ \ \ \ \ \ ({\textbf{artist/bio{\char38}}}\ artist\ bio\ discog)\\
\ \ \ \ \ \ \ \ \ {\textbf{{\textit{;; fetch the artist images}}}}\\
\ \ \ \ \ \ \ \ \ ({\textbf{artist/image{\char38}}}\ artist))))\\
\end{tabular*}
}}
\setlength{\tabcolsep}{\oldpretabcolsep}
}\end{flushleft}
}}
\noindent The \texttt{artist/image{\char38}} subcomponent of \texttt{artist{\char38}} calls FMA and EchoNest in paralle to find an artist image. As soon as one server responds,
the other request is aborted using a \texttt{trap{\char38}} statement with \texttt{exit{\char38}}
triggered when the \texttt{img} local signal is received. If no image is found, the currently displayed image is hidden:

{\footnotesize{\begin{flushleft}
\newdimen\oldpretabcolsep
{\setlength{\oldpretabcolsep}{\tabcolsep}
\addtolength{\tabcolsep}{-\tabcolsep}
{\noindent \texttt{\begin{tabular*}{0.90\linewidth}{l@{\extracolsep{\fill}}}
({\textbf{define}}\ ({\textbf{artist/image{\char38}}}\ artist)\\
\ \ \ {\textbf{{\textit{;; find the first image out of two services}}}}\\
\ \ \ ({\textbf{let}}\ ((el\ (dom-get-element-by-id\ "hophifi-internet-artist-image")))\\
\ \ \ \ \ \ ({\textbf{local{\char38}}}\ ((img\ ({\textbf{instantiate}}{\textbf{::HipHopEvent}}\ (name\ "image"))))\\
\ \ \ \ \ \ \ \ \ {\textbf{{\textit{;; try to find one image on two different servers}}}}\\
\ \ \ \ \ \ \ \ \ {\textbf{{\textit{;; abort the pending request as soon as one returns}}}}\\
\ \ \ \ \ \ \ \ \ ({\textbf{trap{\char38}}}\ (done)\\
\ \ \ \ \ \ \ \ \ \ \ \ ({\textbf{par{\char38}}}\\
\ \ \ \ \ \ \ \ \ \ \ \ \ \ \ ({\textbf{seq{\char38}}}\ \\
\ \ \ \ \ \ \ \ \ \ \ \ \ \ \ \ \ \ ({\textbf{with-hop{\char38}}}\ ({\textbf{{\char36}}}hopfm/artist/image\ ({\textbf{val{\char38}}}\ artist))\ img)\\
\ \ \ \ \ \ \ \ \ \ \ \ \ \ \ \ \ \ ({\textbf{if{\char38}}}\ ({\textbf{now{\char38}}}\ img)\ ({\textbf{exit{\char38}}}\ done)))\\
\ \ \ \ \ \ \ \ \ \ \ \ \ \ \ ({\textbf{seq{\char38}}}\\
\ \ \ \ \ \ \ \ \ \ \ \ \ \ \ \ \ \ ({\textbf{with-hop{\char38}}}\ ({\textbf{{\char36}}}hopfm/artist/image/echonest\ ({\textbf{val{\char38}}}\ artist))\\
\ \ \ \ \ \ \ \ \ \ \ \ \ \ \ \ \ \ \ \ \ img)\\
\ \ \ \ \ \ \ \ \ \ \ \ \ \ \ \ \ \ ({\textbf{if{\char38}}}\ ({\textbf{now{\char38}}}\ img)\ ({\textbf{exit{\char38}}}\ done)))))\\
\ \ \ \ \ \ \ \ \ ({\textbf{if{\char38}}}\ ({\textbf{now{\char38}}}\ img)\\
\ \ \ \ \ \ \ \ \ \ \ \ {\textbf{{\textit{;; update the GUI with the new image}}}}\\
\ \ \ \ \ \ \ \ \ \ \ \ ({\textbf{atom{\char38}}}\\
\ \ \ \ \ \ \ \ \ \ \ \ \ \ \ (node-style-set!\ el\ :visibility\ "visible")\\
\ \ \ \ \ \ \ \ \ \ \ \ \ \ \ (set!\ el.src\ ({\textbf{val{\char38}}}\ img)))\\
\ \ \ \ \ \ \ \ \ \ \ \ {\textbf{{\textit{;; no image was found, hides the current one}}}}\\
\ \ \ \ \ \ \ \ \ \ \ \ ({\textbf{atom{\char38}}}\ (node-style-set!\ el\ :visibility\ "hidden"))))))\\
\\
\end{tabular*}
}}
\setlength{\tabcolsep}{\oldpretabcolsep}
}\end{flushleft}
}}
\noindent 
Note the architectural power of nested preemption.  In
\texttt{artist{\char38}} above, when the {\texttt{artist}} signal
event is received, the \texttt{every{\char38}} preemption loop kills
is body, aborting \texttt{artist/image{\char38}} and by transitivity
its spawned {\texttt{with-hop{\char38}}} requests that might still be
pending. Furthermore, the enclosing
\texttt{until{\char38}\ musicstate} statement of
\texttt{hopfm{\char38}} has an even greater preemption power since it kill
all activities in the program: external preemptions dominate internal ones.

The other HopFM HipHop components are omitted here because their
implementation is similar.

\section{Related Work}
\label{Related works}

We presented a  preliminary version of HipHop at the Plastic'11 workshop
\cite{bns:plastic11}. While the core language has been kept stable, the
integration with HOP has been entirely redesigned and the former
version should be considered as obsolete. 

Orc \cite{KitchinCookMisra2006a} addresses the service coordination
issue by proposing a combinator-based process calculus The temporal
algebra of HipHop is richer than that of Orc. However HipHop does not
yet offer the flexibility of the Orc data-stream pipeline $f>x>g$
operator for large-scale data processing.  Flapjax
\cite{Meyerovich:2009:FPL} provided a unified framework for
client-side event-based programming, based on implicit control defined
by data streams instead of explicit control in Hiphop.  Jolie
\cite{DBLP:journals/entcs/MontesiGLZ07} is a framework to write and
orchestrate Web Services using a service-oriented programming language
inspired by the $\pi$-calculus. However, contrarily to HipHop, Jolie is limited to
server-only orchestration.

\section{Conclusion}
\label{Conclusion}

We have presented HipHop, a new domain-specific synchronous language  
geared to the orchestration of services and user intreaction
within Hop on server and client sides. HipHop deals with logical
events exposed by Hop. Its statements are imported from
Esterel. They are based on temporal sequentiality, concurrency and
preemption, which make it possible to replace the
traditional asynchronous thread~/ event-handler spaghetti \cite{Lee2006} 
by a well-understood 
synchronous programming style imported from embedded systems
programming. The reflexivity of Hop makes it possible to build HipHop programs,
ship them to clients or other servers, and run them. 

With our LastFM example, we have sketched how to orchestrate Web
services and GUI events with HipHop, using its reactive statements as
key architectural tools.

Our current implementation is an interpreter directly
based on Esterel's constructive semantics
\cite{berry2000foundations}. More efficient implementations will
certainly be needed for large-scale applications, see \cite{CompilingEsterel}.
\medskip
\paragraph{\bf Acknowledgements:} we thank Cyprien Nicolas, 
who implemented HipHop in Hop and participated in the PLASTIC'11 first paper about HipHop.

\begin{small}
\bibliographystyle{abbrv} 
\bibliography{hiphop.UTF-8}
\end{small}

\end{document}